# Ultracold atoms in superlattices as quantum simulators for a spin ordering model and phenomena

Godfrey E. Akpojotor

Department of Physics, Delta State University, Abraka 331001, Nigeria

This research is partially funded by Kusmus Communications, ICBR and AFAHOSITECH. Part of this research work was done at the Max Planck Institute for Physics of Complex Systems, Dresden, Germany.

**Abstract**

Cold atoms in optical lattices is the application of two formerly distinct aspects of physics: quantum gases from atomic physics and laser theory from quantum optics. Its use to simulate quantum phenomena and models in condensed matter physics is a growing field. The major goal is to use cold fermonic atoms in these superlattices for the simulations. We present here a theoretical proposal for simulating a spin ordering model using fermions. We demonstrate superexchange interaction in the double well and resonating valence bond (RVB) states in kagome lattice which is important for understanding the $CuO_2$ plane of the superconducting cuprates and other magnetic frustrated materials.

## 1. Introduction

Ultracold atoms in optical lattices have become a growing field for studying quantum strongly correlated systems which exhibit some of the most intriguing phenomena in condensed matter physics such as phase transitions in high temperature superconductivity and spin ordering in magnetism (Georges 2004, Bloch et al. 2008). This increasing interest in the superlattices emanates from the growing advancement in techniques to prepare, manipulate and detect strongly correlated states in them. The emerging artificial quantum crystals are then being used to simulate quantum phenomena and models. The standard Hubbard model is universally considered the simplest minimal description of the strongly correlated systems with its kinetic part expected to account for the itinerancy of the carriers while its Coulombic interaction represents the localization of the carriers. Inspite of this simplicity, the properties of the Hubbard model have been well determined only in the one dimension ($d = 1$) limit where an exact Bethe ansatz solution was obtained in 1968 (Lieb and Wu 1968) and in the infinite dimension ($d = \infty$) limit where an exact solution has been obtained by Dynamical Mean Field theory (DMFT) in 1989 (Metznar and Vollhardt 1989). Thus there is no general consensus for the properties at finite dimensions which is the realm for the real materials. It is pertinent to remark that the model has been investigated with all available theoretical tools that is hoped to be capable of extracting relevant information from the model, as can be found in its rich and vast literature (see Akpojotor (2008a) and references therein for more details). One of these tools is to simulate the model with classical computer but this tool has not been successful because of the general problem of the inability to simulate quantum many body problems computationally due to exponential growth of the required space and time resources with the number of particles which often makes the simulation intractable (J¨ordan et al. 2008). Therefore, a new tool to investigate the Hubbard model was created when Greiner et al. (2002) experimentally demonstrated the superfluid to Mott insulator (SF-MI) transition using cold bosonic atoms in optical lattice which was theoretically proposed for the Bose-Hubbard model by Jaksch et al. (1998).

The SF-MI transition is similar to the metal to Mott insulator transition predicted in the standard Hubbard model which can only be achieved by using fermions since strongly interacting electrons are the carriers responsible for this transition as well as the properties and behaviour of other strongly correlated systems. Therefore there has been a race to use cold fermions in place of the cold bosons used by Greiner et al. (2002) to achieve superfluid to Mott insulator transition. The first seminal work in this race is the observation of the Mott insulator with cold fermonic atoms by J¨ordan et al. (2008). There have been a number of other proposals of using cold fermions in superlattices to simulate quantum phenomena. The purpose of this current study is to give a theoretical proposal on how to simulate a spin ordering model using cold fermions in optical lattices. The plan of the study is as follows. In the next section, we will describe the physics of trapping atoms in optical lattices. Then we will demonstrate how a spin ordering Hamiltonian can be used to achieve superexchange interaction in a double well (DW). We will then go beyond the superexchane interactions in a double well to obtain the resonating valence bond (RVB) states in a kagome lattice which is important in understanding magnetic frustrated systems such as the $CuO_2$ plane believed to be the key feature of the superconducting cuprates (Moessner and Sondhi 2001).

**2. Trapping atoms in optical lattices**

An optical lattice is able to trap atoms because the electric fields of the lasers will induced electric dipole moment in the atom which then interacts with the same electric field of the lasers to produce an effective potential that traps the atom in the optical lattice. The Electric field, E of the standing light wave can be expressed as a product of static spatial and oscillating time-dependent part (Grimm et al. 1999)

$$E(r,t) = \hat{e}E(r)\exp(i\omega t) \tag{2.1}$$

and the induced dipole moment, p oscillating at a driving frequency ω, is

$$p(r,t) = \hat{e}p(r)\exp(i\omega t), \tag{2.2}$$

where E(r) and p(r) are amplitudes and $\hat{e}$ is a unit polarization vector.

The amplitude of the dipole moment is simply related to the field amplitude by

$$p(r) = \alpha(\omega)E(r) \tag{2.3}$$

where $\alpha(\omega)$ is the complex polarizability which depends on the driving frequency.

The field intensity I is related to the field amplitude as

$$I = \varepsilon_0 c |E|^2. \tag{2.4}$$

Taking into account Eq. (2.1) – (2.4), the interaction potential of the induced dipole moment and the driving electric field can be expressed as

$$V_{dip}(r) = -\frac{1}{2\varepsilon_0 c}\text{Re}(\alpha)I(r). \tag{2.5}$$

The interaction between the induced dipole moment and the driving electric field modifies the energy of the atom which is depicted by the electronic transition within the atom. The difference between the frequency of this transition, $\omega_0$ and the frequency of the laser is called detuning $\Delta_t$:

$$\Delta_t = \omega - \omega_0. \tag{2.6}$$

If the laser frequency is less than transition frequency (i.e. $\Delta_t < 0$), the atoms are attracted to potential minima, that is region with maximum electric field intensity and this is known as red detuning while if the laser frequency is greater than the transition frequency (i.e. $\Delta_t > 0$), the atoms are attracted to potential maxima, that is region with minimum field intensity and this is known as blue detuning. Therefore the strength of the optical potential confining the atoms can be increased by tuning the laser intensity, though the atoms can be trapped in either the bright or dark regions of the optical lattice by both types of detuning.

The behaviour of atoms in an optical lattice depends whether they are fermions or bosons. The fermions have half-integer spins and consequently obey the Pauli exclusion principle which states that no two identical fermions can be in the same quantum state at the same time. The physical implication is that fermionic systems will have many energetic particles flying around even as the temperatures goes down to zero as only one particle can occupy the lowest energy. In general, fermions are governed by the Fermi-Dirac distribution (FDD). The bosons, however, have zero or integer spins and consequently do not obey the Pauli exclusion principle. The rules governing the behaviour of photon which is the commonest boson were first given by Satyendra Nath Bose in 1924. Excited by this work, Einstein in the same year extended the rules to other bosons and thereby gave birth to the Bose-Einstein distribution (BED). While doing this, Einstein found that not only is it possible for two bosons to share the same quantum state at the same time, but that they actually prefer doing so. He therefore predicted that when the temperature goes down, almost all the particles in a bosonic system would congregate in the ground state even at a finite temperature. It is this physical state that is called Bose-Einstein condensation (BEC). The Einstein's prediction, however, was considered a mathematical artifact for sometime until Fritz London in 1938 while investigating superfluid liquid helium realized that the phase transition could be accounted for in terms of BEC. This analysis however, suffered a major setback because the helium atoms in the liquid interacted quite strongly. This was why scientists had to move ahead in search of BEC in less complicated systems that would be close to the free boson gas model. Fortunately, the breakthrough came in 1995 when the first BEC was observed in rubidium atoms and this was followed by similar observations in some other cold alkali atoms such as those of lithium and sodium (Anderson et al. 1995; Cornell and Wieman 2002; Hall 2003 and Akpojotor and Ojobor 2008). The achievement of the BEC which won the 2001 Nobel Prize in Physics relied heavily on the then newly developed ability to trap and cool atoms with lasers which was recognized by the Nobel Foundation for the 1997 Nobel Prize in Physics (Metcalf and van der Straten 1999).

The general belief currently is that almost any kind of atom can be trapped in an optical lattice, but alkali atoms are mostly used due to their single valence electron which simplifies description of their behavior in optical lattices. Further, the classification of an atomic isotope to be bosonic or fermionic depends on the number of its constituents: protons, neutrons and electrons. If its number is even, total spin of atom is integer, and the atom is boson while if it is odd, total spin is half integer and the atom is fermion. As bosons the following isotopes are most commonly used: $_{37}Rb^{87}$, $_{11}Na^{23}$, $_{19}K^{39}$, $_{55}Cs^{133}$; and as fermions the commonly used isotopes are: $_{19}K^{40}$, $_{3}Li^{6}$, $_{38}Sr^{87}$.

### 3. Simulations with cold atoms in double wells superlattices

To simulate a physical phenomena or model involves mapping it into alternative physical systems that may be simpler and can easily be manipulated and controlled yet it is described by the same mathematics. The double well is the simplest experimental set up of optical lattices to simulate physical phenomena and models because the system can be completely controlled and measured in an arbitrary two-spin basis by dynamically changing the lattice parameters (Rey et al. 2007). On the theoretical side, the DW can be considered as two localized spatial modes separated by a barrier and consequently be investigated as a two-mode approximation (Jaksch et al. 1998; Akpojotor and Li 2008). It is therefore pertinent to start the simulation with cold atoms in a DW.

The DW is a 1D optical lattice in which the transverse directions are in strong confinement and thus the motions of an atom in these directions are frozen out (Akpojotor and Li 2008). To create the double well superlattice which is simply superimposing one lattice on another, we start with a standing wave of period d and dept $V_1$ (long lattice) and then superimpose on it a counter propagating standing wave with period d/2 and dept $V_2$ (short lattice) as shown in Figure 1a. The resulting superlattice is a 1D symmetric DW (see Figures 1b and 1c) which can be tilted to obtain asymmetric DW as shown in Figure. 1d.

The potential seen by the atom in the superlattice DW is (Akpojotor and Li 2008)

$$V(x) = V_1(x) + V_2 \cos^2(2\pi x/d) \qquad (3.1)$$

Therefore, this potential can be manipulated and controlled by varying the depths of the short and long lattices. For example, by increasing the lattice depth of long-lattice $V_1$, we could reach from superfluid to Mott-insulator regime, which is convenient for studying the few particles phenomena in a local double-well cell. And the barrier height of the double-well is controlled by the lattice depth of short-lattice, $V_2$. The effective double-well is reached if $V_1 > 4V_2$.

For an atom of mass m trapped in any of the wells corresponding to a filling factor of ½, it will undergo a Josephson oscillation with a frequency of

$$\omega = \frac{\pi}{d}\sqrt{\frac{16V_2^2 - V_1^2}{2mV_2}} \quad (3.2)$$

which obviously depends on not only the lattice depths $V_1$ and $V_2$, but also on the lattice spacing d. Usually, the small lattice spacing d is preferred as it leads to a large frequency though this could also be restricted by changing the ratio $V_1/4V_2$. This preference also leads to the use of the recoil energy of the short lattice as the unit of the depths of the optical lattice:

$$E_r = \frac{\hbar^2}{2m\lambda^2} \quad (3.3)$$

where λ is the wave length of the short lattice.

The description so far has been for a 1D symmetric DW (Fig 1c). As discussed above, it can be tilted to obtain asymmetric DW (Fig 1d). The potential bias or the tilt Δ of the DW is introduced by changing the relative phase of the two potentials (i.e. short and long lattices) and this can be realized by applying a magnetic field gradient. Consequently, tuning this field gradient gives the potential difference between the two potential minima of the DW (Trotzky et al. 2008). We can realize the adiabatic and diabatic operations on the tilt of the DW by controlling the increasing speed of the field gradient (Sebby-Strabley et al. 2006).

The starting Hamiltonian for the DW is the two-site version of the Hubbard model (Jaksch et al. 1998; Trotzky et al. 2008 and Akpojotor and Li 2008; 2009):

$$H_H = \sum_{\sigma=\uparrow\downarrow} -J(a^+_{\sigma L}a_{\bar{\sigma} R} + a^+_{\sigma R}a_{\bar{\sigma} L}) - \frac{1}{2}\Delta(n_{\uparrow L} - n_{\downarrow R}) + U(n_{\uparrow L}n_{\downarrow L} + n_{\uparrow R}n_{\downarrow R}) \quad (3.4)$$

where $a^+_{\sigma L,R}(a_{\bar{\sigma} R,L})$ is the creation operator (annihilation operator) for an atom with spin $a(\bar{a}) = \uparrow(\downarrow), \downarrow(\uparrow)$, $n_{a,L,R}$ is the corresponding number operator, $J$ (both J and t are used in the literature though the cold matter community seems to prefer J) describes the tunneling rate between the two wells, Δ is the potential bias for the double-well and $U$ is the two-body interaction when two atoms occupy the same site.

Eq.(3.4) is known as the Bose-Hubbard model which for Δ = 0 has a one to one correspondence with the standard Hubbard model with the former being applicable to systems with bosons as the carriers and the latter to systems with fermions as the carriers

One general consensus, however, is that for spin ordering phenomena and models, there is need to go beyond the Hubbard model either for bosons or fermions (Amadon and Hirsch 1997; Lewenstein and Sanpera 2008; Trotzky et al. 2008; Akpojotor 2008a; Akpojotor and Li 2008; 2009; Liang et al. 2008). For as already explained, achieving the SF-MI transition in the Bose-Hubbard model is simply by varying the potential dept: decreasing the lattice dept will increase the hopping rate so that J dominates while increasing the lattice potential dept will enhance the on-site interaction so that U dominates (Greiner et al. 2002). Introducing a bias potential in the well helps to manipulate the spin ordering of the atoms with very little or no interaction as demonstrated by Trotzky et al. (2008). It follows then that to get nearer to the condensed matter scenario wherein the spin ordering is induced by the interacting spins, the manipulation via bias potential needs to be replaced by interaction mechanism. In otherwords, we make Δ = 0 to achieve symmetric wells and then introduce appropriate interaction via appropriate laser manipulation to achieve the ordering. Such an extended version of Eq.(3.4) has been suggested in previous studies (Akpojotor and Li, 2008; 2009) by including both the nearest neighbor (NN) direct exchange, V and superexchange, $J_{ex}$ interactions (Trotzky et al. 2008) leading to the J-U-V-$J_{ex}$ model:

$$H = -J\left[\sum_{\sigma=\uparrow\downarrow} -J(a^+_{\sigma L}a_{\bar{\sigma} R} + a^+_{\sigma R}a_{\bar{\sigma} L})\right] + U(n_{\uparrow L}n_{\downarrow L} + n_{\uparrow R}n_{\downarrow R}) + V(n_{\uparrow L}n_{\downarrow R} + n_{\downarrow L}n_{\uparrow R}) \quad (3.5)$$
$$- \sum_{\sigma,\bar{\sigma}} J_{ex}(a^+_{\sigma L}a^+_{\bar{\sigma} R}a_{\sigma R}a_{\bar{\sigma} L}).$$

The DW can be prepared initially either as spin singlet states or spin triplet states so that the common basis states (with a basis state $|L,R\rangle$ denoting the L = left and R = right wells) allowed by the Pauli exclusion principle are $|\uparrow_L\downarrow_L,0\rangle$, $|0,\uparrow_R\downarrow_R\rangle$, $|\uparrow_L,\downarrow_R\rangle$, $|\downarrow_L,\uparrow_R\rangle$, $|\uparrow_L,\uparrow_R\rangle$, $|\downarrow_L\downarrow_R\rangle$, where the first two are on-site states, the next two are inter-site states and the last two are triplet states (Rey et al. 2007).

Using our highly simplified correlated variational approach (HSCVA) (Akpojotor 2008a), the exact matrix form of Eq. (3.5) is solved for the DW with fermions to obtain the ground state energy for the singlet ($E_s$) and triplet ($E_t$) states respectively

$$E_s = -2\left[\sqrt{\left(\frac{U}{4J} - \frac{V}{4J} - \frac{J_{ex}}{4J}\right)^2 + 1} - \left(\frac{U}{4J} + \frac{V}{4J} + \frac{J_{ex}}{4J}\right)\right] \qquad (3.6)$$

$$E_t = 4\left(\frac{V}{4J} - \frac{J_{ex}}{4J}\right) \qquad (3.7)$$

where U/4J is the on-site interaction strength which determines the response of the kinetic energy of the electrons to the varying on-site Coulombic interaction U, V/4J is the NN inter-site interaction strength which determines the response to the varying NN Coulombic interaction and $J_{ex}$/4J is the NN superexchange interaction strength which determines the response to the varying superexchange interaction $J_{ex}$. All these quantities are physically dimensionless as they are ratios of the same unit. As expected, the ground state energy for the triplet state is double fold degenerate and this emanate from the up-spins and the down-spins.

It has also been shown in our aforementioned previous studies (Akpojotor and Li 2008; 2009) that the transition point, $T_p$ of the ground state energy from $E_s$ (i.e. antiferromagnetic ordering) to $E_t$ (i.e. ferromagnetic ordering) is when

$$\frac{J_{ex}}{4J} > \frac{1}{2}\left[\sqrt{\left(\frac{U}{4J} - \frac{V}{4J}\right)^2 + \frac{1}{2}} - \left(\frac{U}{4J} + \frac{V}{4J}\right)\right]. \qquad (3.8)$$

The $T_P$ could be sharp, meaning that there is complete cross over from the antiferromagnetic phase to the ferromagnetic phase so that $E_t$ is never equal to $E_s$ for all values of J/4t. It has been shown (Akpojotor and Li 2009) how this sharp $T_P$ can be used to account for the first experimental demonstration of superexchange interaction with cold atoms in optical lattices (Trotzky et al. 2008). Further, prediction of the theoretical values of the tunneling parameter and the interaction parameters from extracted data from the experiment of the possible dynamic evolution frequencies,

$$\hbar\omega_{1,2} = \frac{\sqrt{16J^2 + U^2} \pm U}{2}, \qquad (3.9)$$

$$\hbar\omega_{3,4} = 2J\left[\sqrt{\left(\frac{U}{4J} - \frac{V}{4J} - \frac{J_{ex}}{4J}\right)^2 + 1} \pm \left(\frac{U}{4J} + \frac{V}{4J} + \frac{J_{ex}}{4J}\right)\right] \qquad (3.10)$$

and $\qquad \hbar\omega_5 = V - J_{ex} \qquad (3.11)$

have been obtained as $J = \frac{1}{2}\hbar\sqrt{\omega_1\omega_2}$, $U = \hbar(\omega_1 - \omega_2)$, $V = \frac{\hbar}{2}(\omega_3 - \omega_5) - (\omega_1 - \omega_2) + \omega_5$ and $J_{ex} = \frac{\hbar}{2}(\omega_3 - \omega_5) - (\omega_1 - \omega_2) - \omega_5$ (Akpojotor and Li 2008).

The other possible case for Eq. (3.8) is for the $T_P$ not to be sharp, meaning there is still antiferromagnetic ordering at the onset of that ferromagnetism (i.e $E_t = E_s$) at certain values of J/4t before completely crossing over to the ferromagnetic phase. This scenario of co-existence is called the mixed state and it can be used to investigate two interesting phenomena in condensed matter physics, superconductivity and magnetically frustrated systems.

**4. Simulations with cold atoms in superlattices beyond the double wells**

To investigate the superconductivity, one has to start from the observation by Lewenstein et al. (2007) that the mixed states describe the resonating valence bond states. Here the electrons are localized to individual atoms and the fluctuations in the charge (density) degree of freedom are strongly suppressed. Therefore the physics is dictated by the remaining spin degree of freedom which interact via superexchange and this can result to the superposition of states in which random pairs of neighbouring pairs of atoms attains zero total spin (Nascimbene et al. 2012). Interestingly, it is this RVB states that was adopted by Anderson (1987) in his proposal that the 2D $CuO_2$ plane which is generally believed to be the key feature to understanding the high $T_c$ superconducting cuprates, can be reduced to a single band pairing problem. This was confirmed by Zhang and Rice (1988) who then showed that the ground state will be a singlet pairing of the Cu and O and that if liberated as a Cooper pair, will lead to superconductivity. We have been able to show the formation of the singlet pairing as the Cooper pair and its propagation within a superexchange interaction (Akpojotor 2008b). Therefore the design and implementation of the formation and propagtion of the Cooper pair using cold atoms in optical lattices can be achieved by improving and extending the study of superexchange interaction in DWs (Trotzky et al. 2008) to a $CuO_2$ plaquette (see Figure 2). This has been recently achieved by Nascimbene et al. (2012) using cold bosons. The use of cold fermions to achieve it experimentally is still an outstanding problem. So here we give the theoretical insight to aid the experimental realization with cold fermions by studying the plaquette as a half filled 4 x 4 square lattice which can also be used for studying other magnetically frustrated systems.

In general, the magnetically frustrated systems such as the kagome lattice can be studied as the mixed state of the t-U-V-J model because numerical results of the spin $1/2$ system of the Kagome lattice suggest that the energy gap between the ground state and the lowest triplet state, if any, is very small (of the order of $J_{ex}/20$) and that this gap is filled with low-lying singlets (Waldtmann et al. 1998). As pointed by Lewenstein et al. (2007), these results suggest that the frustrated systems can be described as the RVB states.

The kagome lattice is a 2D frustrated system composed of corned-shared triangles. Since the triangular geometry is believed to have frustrated magnetic ordering (see Figure 3a), the kagome lattices are believed to exhibit the behaviour of magnetically frustrated materials (Moessner and Sondhi 2001). Frustration here means all the constraints imposed by the Hamiltonian cannot be simultaneously fulfilled (Akpojotor and Akpojotor 2009).

To design the Kagome lattice with cold atoms in optical lattice, it can be mapped into a 4 x 4 square lattice (sites 22, 23, 32 and 33) as shown in Figure 3b. Adopting the method of Nascimbene et al. (2012), we can realize the RVB states using cold fermions by loading the 4 x 4 cluster at half-filling, that is, four atoms per site. It is then easy to see that for Fermi gas there will be a total of 256 singlet states and 240 triplet states while for bosonic atoms there be a total of 136 states. Using the highly simplified correlated variational approach (HSCVA) (Akpojotor 2008a), the t-U-V-J Hamiltonian in Eq. (3.5) for Fermi gas will yield an 11 x 11 matrix which is then solved numerically to obtain the ground state energy at the $T_p$ as the $J_{ex}/4J$ is increased from zero at $U/4J = 3$ and $V/4J = 0$. The results which is depicted in Figure 4 clearly show a mixed state which is the RVB states that emanates from the spin ordering frustration in the Kagome lattice.

**Conclusion**

One challenge in the experimental verification of the superexchange model investigated in this work is to develop a method to cool Fermi gas to temperature below the $J_{ex}$ because if the temperature is not cold enough, thermal fluctuations would destroy the fragile magnetic order present in the ground state (Bloch 2008). This is why we have shown the result for $U/4J = 3$ to allow the verification of our proposal here with even bosons as a first step. For the $U/4J = 3$ means a strong coupling regime which will map the bosons into non-interacting fermions. In this process known as fermionization, the hardcore repulsion mimics the exclusion principle. Then one can encode the spin $1/2$ state as presence of (↑) or absence (↓) of boson at the site. Another method that can be used to verify our study is the Eckardt and Lewenstein (2010) robust implementation of a quantum simulator for the homogeneous $J-J_{ex}$ model with well controlled hole doping, using a sample of ultracold bosonic and fermionic atoms in an optical lattice.

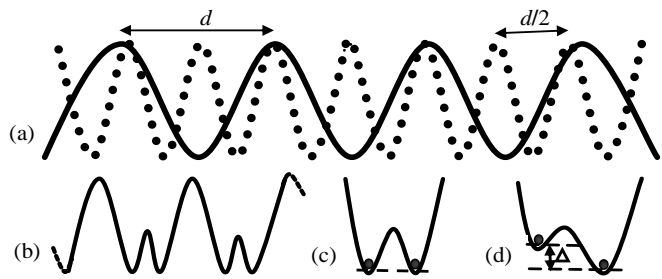

Figure 1 (a) Two standing waves in opposite directions and with periods d and d/2 resulting in (b) a chain of double wells from which we can study (c) a symmetric double well or (d) an asymmetric double well.

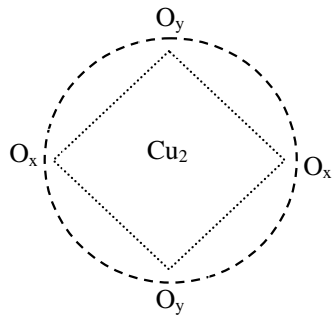

Figure 2. The $CuO_2$ plaquette which is generally believed to be the key feature to understanding the high Tc superconducting cuprates

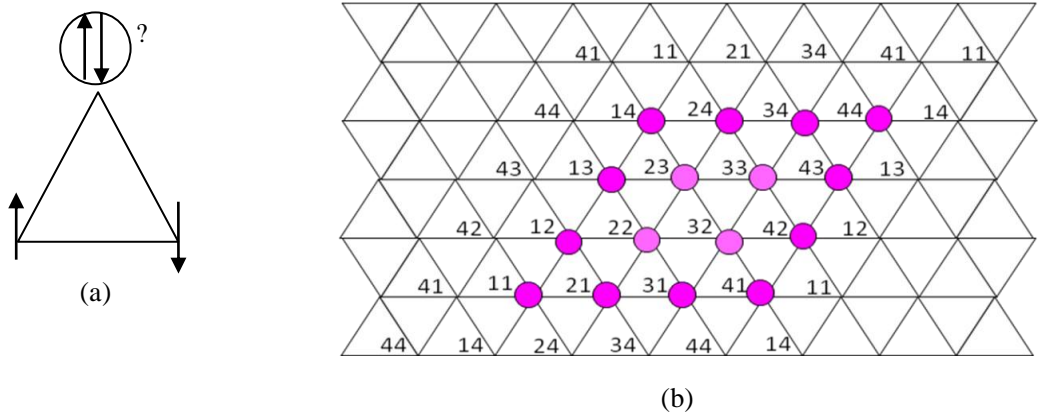

Figure 3 (Colour online): (a) The triangular lattice as a geometrically frustrated spin system (b) The Kagome lattice formed by the triangular lattices with the round purple circles indicating the possible trapping of cold atoms in a 4 x 4 cluster.

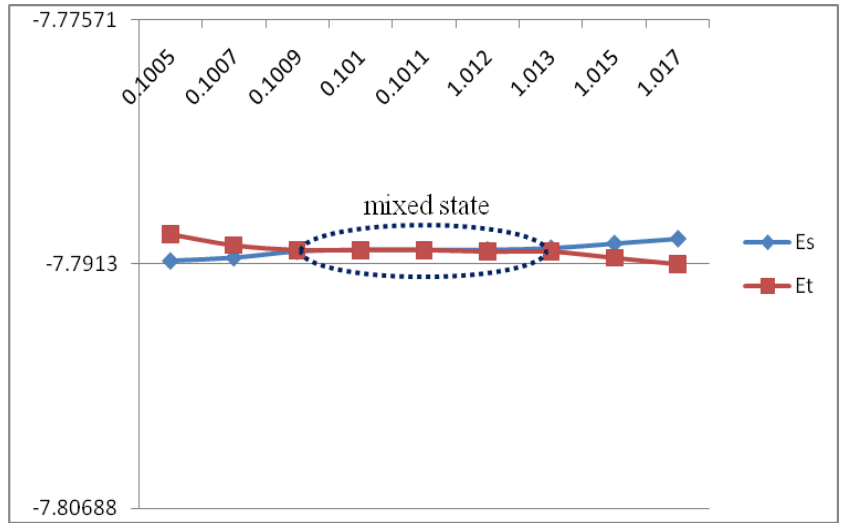

Figure 4 (Colour online): The change of the ground state energy from antiferromagnetic ordering to ferromagnetic ordering at the transition point, $T_p$ as the superexchange interaction, $J_{ex}/4J$ is increased from zero at $U/4J = 3$ and $V/4J = 0$ for a Kagome lattice of 4 x 4 cluster.